\documentclass{aa}

\usepackage{graphicx}
\usepackage{times}

\def\approxgt{\mathrel{\hbox{\rlap{\lower.55ex \hbox {$\sim$}}
        \kern-.3em \raise.4ex \hbox{$>$}}}}
\def\approxlt{\mathrel{\hbox{\rlap{\lower.55ex \hbox {$\sim$}}
        \kern-.3em \raise.4ex \hbox{$<$}}}}

\begin{document}
\thesaurus{6(02.01.2; 11.01.2; 11.05.1; 11.09.1 (NGC~1052); 11.14.1; 13.25.2)}

\title{On the hypothesis of an advection-dominated flow in the core of NGC~1052: new constraints from a BeppoSAX observation}

\author{M. Guainazzi\inst{1}, T. Oosterbroek\inst{2}, L.A. Antonelli\inst{3}, G. Matt\inst{4}}

\institute{
{XMM-Newton SOC, VILSPA, ESA, Apartado 50727, 28080 Madrid, Spain}
\and
{Astrophysics Division, Space Science Department of ESA, ESTEC, Postbus 299, NL-2200 AG Noordwijk, The Netherlands}
\and
{Osservatorio Astronomico di Roma, Via dell'Osservatorio, I-00044 Monteporzio Catone, Italy}
\and
{Dipartimento di Fisica, Universit\`a degli Studi Roma Tre, Via della Vasca Navale 84, I-00046 Roma, Italy}
}
   
\offprints{M.Guainazzi}

\date{Received  ; accepted }

\maketitle

\markboth{M.Guainazzi et al.}{On the hypothesis of an advection-dominated flow in the core of NGC~1052}
   
\begin{abstract}

We report the results of a broadband (0.1--100~keV) X-ray observation of
the nearby elliptical galaxy NGC~1052, performed with the BeppoSAX
observatory. We confirm the presence of a bright
(2--10~keV luminosity $\sim$$4 \times 10^{42}$~erg~s$^{-1}$) and
strongly absorbed ($N_{\rm H} \sim 2 \times 10^{23}$~cm$^{-2}$)
X-ray source. The flatness of the
X-ray spectrum (photon index, $\Gamma$, $\simeq$$1.4$),
the estimated low accretion rate ($\dot{m} \equiv \dot{M}/\dot{M}_{\rm Edd}
\sim 10^{-4}$) and the
radio-to-X-ray spectral energy distribution suggest that this observation
may represent the first direct measurement above 10~keV of
 an accretion-dominated flow in an elliptical galaxy.
\end{abstract}

\keywords{ Accretion --
	   Galaxies:active --
	   Galaxies:elliptical and lenticular --
	   Galaxies:individual:NGC~1052 --
	   Galaxies:nuclei --
  	   X-ray:galaxies}

\section{Introduction}
The discovery of supermassive black holes
($10^6$--$10^9 M_{\rm \odot}$)
in the nuclei of
several nearby galaxies (Magorrian et al. 1998) has raised
the question of why most of them are not
active. A possible solution
is provided by the so called Advection Dominated Accretion Flows
(ADAF; Rees 1982;
Narayan \& Yi 1995; Fabian \& Rees 1995). In this class
of accretion solutions, the accreting gas is so tenuous that it
cannot cool efficiently, the viscous energy is
stored in the protons as thermal energy
and eventually advected onto the nuclear
compact object. The ADAFs are therefore characterized by small
radiative efficiency and accretion rates
($\dot{m} \equiv \dot{M}/\dot{M}_{\rm Edd} < 10^{-1.6}$; Narayan \& Yi
1995; Rees et al. 1982). At low $\dot{m}$
the hard X-ray emission is
mainly due to bremsstrahlung emission from a population of $\sim$$100$~keV
electrons,
and is therefore much harder than typically
observed in Seyfert galaxies (Nandra et al. 1997; Turner et al. 1997;
Matt 2000).

Recently, hard X-ray tails
have been discovered in the ASCA spectra of several elliptical
galaxies (Allen et al 2000).
Photon indices are in the range 0.6--1.5, therefore
remarkably consistent with the expectations of the ADAF scenario.
Unfortunately, the ASCA energy bandpass is
limited to 9--10~keV, and the detection of these hard tails is therefore
difficult and somewhat model-dependent.

One of the most intriguing of these ADAF-candidates is the nearby
($z = 0.0049$) elliptical
galaxy NGC~1052.
It is
a narrow-line
LINER (Heckman 1980), with a compact radio core (diameter
$\simeq$$0.14$~pc) and a radio halo of about 3~kpc diameter.
NGC~1052 is the first type~2
LINER where broad lines in spectropolarimetric measurements
have been discovered
(Barth et al. 1999). The ASCA spectrum (Guainazzi \& Antonelli 1999, G99;
Weaver et al. 1999) is 
extremely flat: a formal fit with a simple power law yields a photon index
$\Gamma \simeq 0.1$. This suggests either high photoelectric
absorption 
(column density $N_{\rm H} \simeq 10^{23}$~cm$^{-2}$) or a
Compton-reflection dominated spectrum (which, of course, would imply an
even more strongly absorbed active nucleus). In the former scenario
the intrinsic spectral index is still very flat ($\Gamma \simeq 1.2$).
This is intriguingly consistent with the ADAF scenario.

The ASCA results prompted a BeppoSAX program of
observations of narrow-line LINERs with broad spectropolarimetric lines,
mainly focused to investigate their hard X-ray spectrum
($\approxgt$10~keV). The results of this experiment on NGC~1052 are
reported in this paper.

\section{Observation and data reduction}

BeppoSAX observed NGC~1052 between 2000
January 11 (19:27 UT) and 13 (08:57 UT). The instruments were operating in
nominal direct modes. Data reduction and analysis
followed standard procedures, as detailed, {\it e.g.}, in Guainazzi et 
al. (1999). Scientific products from the imaging instruments (Low
Energy Concentrator Spectrometer, LECS; Parmar et al. 1997;
Medium Energy Concentrator Spectrometer, MECS; Boella et al. 1997)
were extracted from circular regions around the galaxy X-ray centroid
of radii 2$\arcmin$ and 4$\arcmin$, respectively. Background subtraction
was performed using deep blank sky field exposures, accumulated by the
BeppoSAX Science Data Center (SDC). The background-subtracted net count rates
are $(1.26 \pm 0.09) \times 10^{-2}$~s$^{-1}$ and $(3.71 \pm 0.08)
\times 10^{-2}$~s$^{-1}$, in the LECS (0.1--4~keV) and MECS (1.8--10.5~keV),
respectively, corresponding to exposure times of 25.8 and
63.2~ks. Spectral analysis made use of the latest response matrices
released by the SDC in January 2000.

In what follows: energies are quoted in the source rest frame, and errors
are at 90\% confidence level for one interesting parameter ($\Delta
\chi^2 = 2.71$), unless otherwise specified.

\section{Results}

\subsection{The PDS spectrum ($E>10$~keV)}

\begin{figure}
\hbox{
\includegraphics*[width=6.0cm,height=8.0cm,angle=-90]{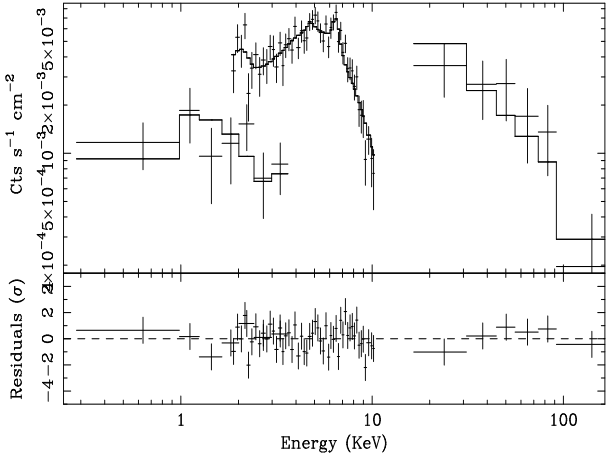}
}
\caption{Spectrum plus best-fit model ({\it upper panel}) and residuals
in units of standard deviations ({\it lower panel}) when the model ``P''
is applied to the broadband BeppoSAX NGC~1052 spectrum}
\label{fig1}
\end{figure}

The 60.0~ks exposure time on NGC~1052 allowed a PDS detection,
with a total 13--200~keV count rate of
$0.22 \pm 0.04$~s$^{-1}$. If we take into account the typical systematic
uncertainties associated with the PDS background subtraction algorithm
(Guainazzi \& Matteuzzi 1997), the detection is at a level higher
then 5$\sigma$ in the 13--90~keV band (see Fig.~\ref{fig1}).
No known hard X-ray bright source is present in the  (1.3$^{\circ}$)$^2$
PDS field of view.
The number of expected sources with flux equal or higher than NGC~1052
is $\approxlt 0.06$ according to the
Cagnoni et al. (1998) logN-logS relation.
A simple power-law fit of the PDS spectrum yields a very flat index:
$\Gamma = 1.0 \pm^{0.7}_{0.3}$.

\subsection{The broadband BeppoSAX Spectrum (0.1--100~keV)}

The source did not show any significant X-ray variability during the
BeppoSAX observation.  The reduced $\chi^2$, when a constant line is fit
to the $\Delta t = 5760$~s light curve are
$\chi^2_{\nu} = 0.76$ and $\chi^2_{\nu} = 0.78$ in the 0.1--2~keV (LECS)
and 2--10~keV (MECS) energy bands, respectively. We will therefore focus in
this Section on the time-averaged spectra only.

A photoelectrically-absorbed single component model provides an
inadequate fit of the broadband (0.1--100~keV)
BeppoSAX spectrum ({\it e.g.}: $\chi^2 = 185.4/80$ degrees
of freedom, dof, if a power-law
model is employed).
On the other hand, a very good fit ($\chi^2_{\nu} \simeq 1.02$--$1.03$)
is obtained with a two-component model, constituted by an absorbed
($N_{\rm H} \simeq$ a few $10^{23}$~cm$^{-2}$) power-law
plus a ``soft excess'' below 2~keV.
The limited statistics
prevents us from unambiguously characterizing the latter component. In the
following
we will discuss, as illustrative examples, models where this soft excess
is described: either with a power-law (``P'' model hereinafter), whose index
is held fixed to that of the high-energy absorbed power-law (thus
modeling reflection of the nuclear continuum, scattered by an
electron plasma - ``warm mirror'' - along our line of sight;
Antonucci \& Miller 1985);
or with thermal emission from a collisionally
ionized, optically thin plasma ({\tt mekal} model in {\sc Xspec};
``M'' model hereinafter).
Model ``P'' also describes a geometry, in which the absorber
only covers a fraction of the line of sight.
The best-fit parameters
are reported in the upper panel of Tab.~\ref{tab1}.
As already suggested by the analysis of the PDS spectrum alone,
the absorbed power-law component is rather flat ($\Gamma \simeq 1.4$).
If the power-law in model ``P'' is substituted by a thermal bremsstrahlung
($\chi^2=83.6/77$~dof), its temperature is $150 \pm^{720}_{100}$~keV.
\begin{table*}
\begin{center}
\begin{tabular}{lcccccc} \hline \hline
Model & $N_{\rm H}$ & $\Gamma$ & $kT$/$f_{\rm s}$ & $E_{\rm c}$ & $EW$  &  $\chi^2$/dof \\
& $10^{23}$~cm$^{-2}$ & & (keV/\%) & (keV) & (eV) & \\ \hline
\multicolumn{7}{l}{BeppoSAX} \\
M & $2.0 \pm^{0.6}_{0.5}$ & $1.39 \pm^{0.25}_{0.18}$ & $>5$ & $6.48 \pm^{0.16}_{0.20}$ & $230 \pm 170$ & 77.7/76 \\
P & $2.1 \pm 0.5$ & $1.45 \pm^{0.20}_{0.16}$ & $20 \pm^{15}_{11}$ & $6.54 \pm^{0.17}_{0.14}$ & $310\pm 70$ & 79.6/77 \\ \hline
\multicolumn{7}{l}{ASCA-ROSAT} \\
M & $0.9 \pm^{0.6}_{0.3}$ & $1.0 \pm^{0.5}_{0.4}$ & $>9$ & $6.35 \pm^{0.07}_{0.06}$ & $220 \pm 110$ & 296.0/309 \\
P & $1.2 \pm 0.2$ & $1.22 \pm 0.16$ & $26 \pm^{15}_9$ & $6.36 \pm 0.06$ & $300\pm 110$ & 297.0/310 \\ \hline
\multicolumn{7}{l}{ASCA-BeppoSAX-ROSAT} \\
M & $2.0 \pm^{0.6}_{0.5}$$^a$ & $1.4 \pm 0.2$ & $>11$ & $6.38 \pm^{0.07}_{0.06}$ & $280 \pm 90$ & 383.0/444 \\
 & $1.2 \pm^{0.3}_{0.2}$$^b$ &  &  &  &  &  \\
P & $1.9 \pm 0.4$$^a$ & $1.33 \pm^{0.09}_{0.08}$ & $22 \pm^4_5$ & $6.40 \pm^{0.06}_{0.07}$ & $300\pm 90$ & 386.0/445 \\ 
 & $1.18 \pm^{0.18}_{0.16}$$^b$ &  &  &  & & \\ \hline
\end{tabular}
\end{center}
\caption{Best-fit parameters and results when the models ``M'' and ``P'' (details in text) are applied to the NGC~1052 broadband spectrum of BeppoSAX ({\it upper panel}), ASCA-ROSAT (after G99; {\it central panel}), and ASCA-BeppoSAX-ROSAT ({\it lower panel}). $f_{\rm s}$ is the scattering fraction (defined as the 2--10~keV flux ratio between the transmitted and the scattered power-law components). $E_{\rm c}$ and $EW$ are the centroid energy and the equivalent width of the emission line, respectively.}

\noindent

$^a$BeppoSAX

$^b$ASCA

\label{tab1}
\end{table*}

The addition of a narrow Gaussian emission line
is required at the 98.9\% confidence level, according to
the F-test, in the ``P'' model ($\Delta \chi^2|_P = 10$ for a
decrease of the degrees of freedom by two), whereas only at the
90.9\% level
in the ``M'' model  ($\Delta \chi^2|_M = 5$).
The Gaussian line centroid energy is consistent, within the statistical
uncertainties, with K$_{\rm \alpha}$ fluorescent emission from neutral
iron. However,
the $EW$ of the iron line
system is too large to be produced in transmission by the same cold matter,
which is responsible for the attenuation of the X-ray continuum
(which would imply $EW \simeq$130~eV for a spherical
distribution of matter;
Leahy \& Creighton 1993). No iron emission line is expected from
an ADAF. The slight difference in $EW$ between models
``P'' and ``M'' (in the latter the iron line profile
is partly accounted by the emission of the thermal plasma)
may suggest a multi-component
structure of the iron line, which is unresolved by the MECS.
Ionized iron lines could be also produced
by the ``warm mirror''
(Netzer \& Turner 1997).
We have therefore repeated the fit in the ``P'' scenario,
assuming that the iron emission
actually consists of two components: one neutral
($E_{\rm c} = 6.4$~keV) and one He-like ($E_{\rm c} = 6.7$~keV).
The fit is of comparable quality ($\chi^2 = 79.8/77$~dof), with:
$EW$(6.4~keV)~$=170\pm^{170}_{150}$~eV; 
$EW$(6.7~keV)~$=180\pm^{150}_{160}$~eV.
The soft excess continuum flux at 6~keV is about 1/3 of that of
the transmitted component. Therefore, the EW of the ionized iron line
against its proper continuum would be of the correct order of magnitude
if produced in a ``warm mirror'' (Matt et al. 1996). The
neutral component $EW$ is now consistent with being produced
in transmission by the same matter covering the active nucleus, if its
covering factor is large.

A hard X-ray continuum could be in principle due to Compton reprocessing
of the nuclear continuum, by either the accretion disc (George \& Fabian
1991; Matt et al. 1992) or the molecular
torus encompassing the active nucleus (Ghisellini et al. 1994;
Krolik et al. 1994). This scenario
does not, however, match our data.
A fit, where the absorbed high-energy component is a bare face-on
Compton-reflection (model {\tt pexriv} in {\sc Xspec};
Magdziarz \& Zdziarski 1995) is statistically unacceptable ($\chi^2 =
112.6/76$~dof).
The addition of a Compton-reflection
component to the ``P'' model (where only the relative
normalization between the reflected and the direct component, $R$,
and the intrinsic power-law cut-off energy
are left free parameters in the fit; an inclination
angle of $30^{\circ}$ and solar abundances are assumed)
does not significantly improve the fit ($\chi^2 = 79.1/75$~dof).
The 90\% upper limit for two interesting parameter
on R is 0.6 (for $\Gamma < 1.65$). These results
allow us to rule out one class of models, which adequately fit
the ASCA-ROSAT spectra, hence favoring
the G99 transmission scenario.

The observed flux in the 0.5--2~keV (2--10~keV) energy band is 0.4
(4.0)$\times 10^{-12}$~erg~cm$^{-2}$~s$^{-1}$. This corresponds
to a luminosity of 0.4 (4.2)$\times 10^{41}$~erg~s$^{-1}$.

\subsection{Comparison with ROSAT and ASCA data}

In the central panel of
Tab.~\ref{tab1}, the best-fit parameters are reported, when the
``M'' and ``P'' models are applied to the ASCA and ROSAT spectra
of NGC1052 (see the Tab.~1 in G99). NGC~1052
was comparatively bright during the August 1996 ASCA
(2--10~keV flux $\simeq$$3.5 \times 10^{-11}$~erg~cm$^{-2}$~s$^{-1}$)
and the January 2000 BeppoSAX observations.
The only spectral parameter showing a
significant difference is
the absorbing column density, which
was about a factor of 2 higher in the later BeppoSAX observation. The spectral
indices measured by BeppoSAX tend also to be slightly softer, but
still consistent with the ASCA-ROSAT measurements within the statistical
uncertainties.

\begin{figure}
\hbox{
\includegraphics*[width=6.0cm,height=8.0cm,angle=-90]{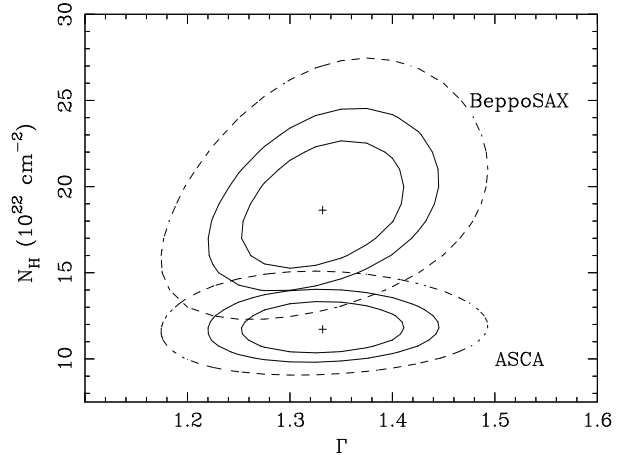}
}
\caption{Spectral index versus column density iso-$\chi^2$ contour plots,
when the ``"P" model is simultaneously applied to the ASCA, BeppoSAX
and ROSAT spectra. The column density is
left free to vary independently in the BeppoSAX and ASCA/ROSAT
models ({\it labels}). The contours correspond to the 68\%, 90\% ({\it
solid lines}) and 99\% ({\it dashed line}) confidence levels for
two interesting parameters.}
\label{fig4}
\end{figure}

We have performed a simultaneous fit of the ROSAT, ASCA and
BeppoSAX spectra, to check whether the improved statistics allows us
to distinguish between the ``M'' and ``P'' models. In these fits
only the column density absorbing the primary nuclear continuum
has been allowed to vary independently in the BeppoSAX and ASCA-ROSAT models
(ROSAT spectra are basically insensitive to column densities of the
order of $10^{23}$~cm$^{-2}$).
Normalization constants have been included as free parameters in the models,
to account for the different fluxes measured in the three observations.
The results are reported in the
lower panel of Tab.~\ref{tab1}. The two models yield comparably
good fits ($\chi^2|_P = 386.0/445$~dof;
$\chi^2|_M = 383.0/444$~dof). Better data quality is needed
to resolve this issue. The spectral index is indeed
much better constrained
than by BeppoSAX data alone, and still very flat (see Fig.~\ref{fig4}).

\section{Discussion}

NGC~1052 is the first ``type~2'' LINER, where broad lines were discovered
in spectropolarimetric measurement (Barth et al. 1999).
One may therefore expect that its
X-ray spectrum resembles that of Seyfert~2 galaxies, where the
luminous nuclear emission is photoelectrically absorbed by matter with
a substantial column density. The observation
performed with BeppoSAX
demonstrates that
this is indeed the case, thanks to the first measure of the broadband
(0.1--100~keV) X-ray spectrum of this object.

\begin{figure}
\hbox{
\includegraphics*[width=6.0cm,height=8.0cm,angle=-90]{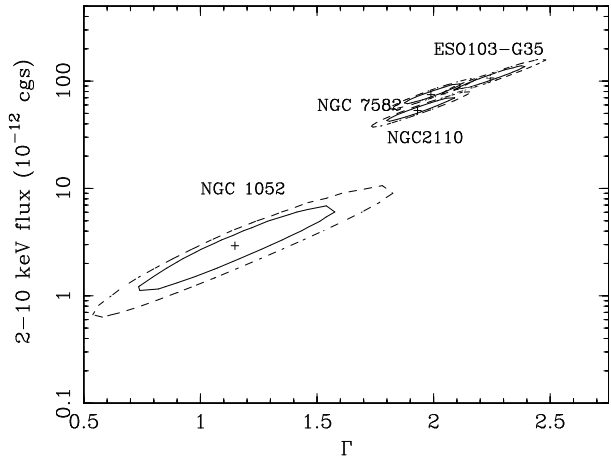}
}
\caption{Spectral index versus extrapolated 2--10~keV flux iso-$\chi^2$
contours if a simple
power-law model is applied to the PDS spectrum of NGC~1052 and of the
``flat'' Seyfert~2s of Turner et al. (1997). The levels are 68\% ({\it
solid line}) and 90\% ({\it dashed line}) for two interesting parameters}
\label{fig2}
\end{figure}

Even more intriguing is the nature of the accretion flow
in this AGN. The nuclear spectrum is rather flat ($\Gamma \simeq
1.4$), as the ASCA observation
had already suggested (G99; Weaver et al. 1999). The best-fit value
is inconsistent
at 2$\sigma$ level with the distribution of spectral indices observed in
Seyfert~1s ($\langle \Gamma_{\rm Sy1} \rangle = 1.87$;
$\sigma_{\rm Sy1} = 0.22$;
Nandra et al. 1997; Reynolds 1997) and radio-quiet quasars
($\langle \Gamma_{\rm RQQ} \rangle = 1.93$; $\sigma_{\rm RQQ} = 0.22$;
Reeves \& Turner 2000).
It is closer to the distribution of spectral indices in radio-loud
quasars ($\langle \Gamma_{\rm RLQ} \rangle = 1.6$;
$\sigma_{\rm RLQ} = 0.15$;
Reeves \& Turner 2000). However, NGC~1052 radio-loudness, according to
the Wilkes \& Elvis (1987) definition, is 1.1, therefore pointing to
a radio-quiet or, at most, a borderline object. There are some
Seyfert~2s, whose ASCA spectral indices (Turner et al. 1997)
are as flat as those measured in NGC~1052. Recently, Malaguti et al. (1999)
pointed out that at least in one case (NGC~2110)
this could be due to an incorrect
modeling of a complex photoelectric absorber in the ASCA bandpass. To verify
this hypothesis, we have extracted from the BeppoSAX archive the PDS spectra
of the ``flat'' Seyfert~2s of the Turner et al. (1997) sample.
In Fig.~\ref{fig2}
we compare the $\Gamma$ versus 2--10~keV (extrapolated) flux iso-$\chi^2$
contours for NGC~1052
and the ``flat'' Seyfert~2s (the results for NGC~5252 are not shown, because
the PDS detection is too weak). In all cases the intrinsic spectral index as
measured by the PDS is significantly steeper than in ASCA. This
suggests that the nature of the accretion flow in the nucleus of NGC~1052
is qualitatively different from that normally at work in
bright nearby AGN.

The NGC~1052 high-energy spectrum is can be also
well approximated by a thermal
bremsstrahlung with $kT \simeq 150$~keV, in
agreement with the expectations of the ADAF scenario.
No estimate of the
nuclear black hole mass is
available for NGC~1052. If we use the bulge B magnitude (Ho et al. 1997)
versus black hole mass relation of Magorrian et al. (1998), we obtain
$M_{\rm BH} \sim 10^{8 \pm 1} M_{\rm \odot}$. The
bolometric luminosity of
the ADAF component extrapolated from the BeppoSAX measurement is $\simeq$$
5 \times 10^{42}$~erg~s$^{-1}$, suggesting an accretion rate $\dot{m}
\sim 10^{-4 \pm 1}$. This value is significantly lower than
the critical $\dot{m}$,
below which the onset of the ADAF regime would occur (Narayan \& Yi 1995).

\begin{figure}
\hbox{
\includegraphics*[width=8.0cm,height=8.0cm]{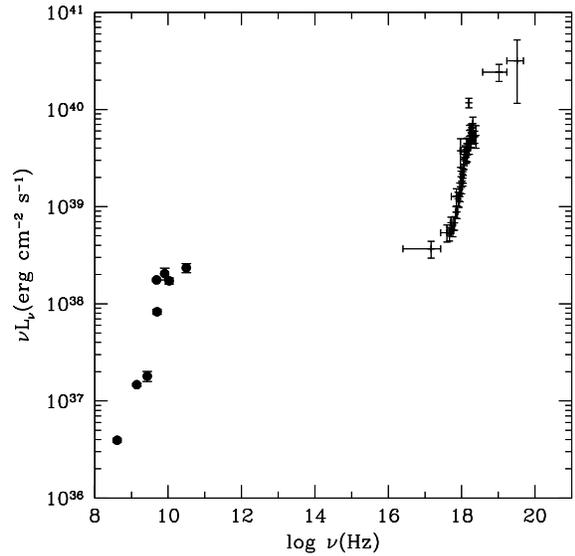}
}
\caption{Radio-to-X-ray SED of NGC~1052}
\label{fig3}
\end{figure}

Critical diagnostic tools for the existence of an ADAF are the radio-to-X-ray
Spectral Energy Distribution (SED) and luminosity
ratio (Di Matteo et al. 2000). In
Fig.~\ref{fig3} we show the NGC~1052 radio/X-ray SED. The radio
data points are taken from the NED archive.
We observe a good qualitative agreement between the shape of the SED
in NGC~1052 and in the ``candidate-ADAF'' elliptical galaxies
of the Allen et al. (2000) sample.
In Tab.~\ref{tab2} we compare the ratios between the powers at
6~keV and 5~GHz in these galaxies and in NGC~1052.
In the latter this ratio is well within the (rather narrow)
\begin{table}
\begin{center}
\begin{tabular}{cc} \hline \hline
Source & $\log(P_{\rm 6 \ keV}/P_{\rm 5 \ GHz})$ \\ \hline
NGC~1052 & 2.5 \\
NGC~1399 & 2.0 \\
NGC~4472 & 2.1 \\
NGC~4486 & 2.8 \\
NGC~4696 & 3.0 \\
NGC~4649 & 2.7 \\ \hline
\end{tabular}
\end{center}

\caption{X-ray to radio power ratios in NGC~1052 and the candidate
ADAF elliptical galaxies of Allen et al. (2000). The data for these
objects are taken from Di Matteo et al. (2000)}
\vspace{-0.5cm}
\label{tab2}
\end{table}
range observed in the former objects, confirming the qualitative similarity.
For comparison, the value of this ratio is typically $\approxgt 10^5$
in radio-quiet quasars (Elvis et al. 1994)

It is also interesting to compare the X-ray with the IR photometry
of Becklin et al. (1982). Their 1.2$\mu$m flux in the innermost 2$\arcsec$
($\simeq$300~pc) - 26~mJy - corresponds to a $L(1.25 \ \mu m)/L(5 \ keV)$
ratio of about 15. Such a high value is typical of Seyfert~2 galaxies
(Mass-Hesse et al. 1995). The IR emission is therefore likely to be
dominated by heated dust, totally masking the ADAF IR intrinsic
emission and therefore preventing us from checking the presence of a wind
in the ADAF scenario (Di Matteo et al. 2000).

The luminosity of the broad component of H$_{\rm \beta}$ is
$\sim$$2 \times 10^{41}$~erg~s$^{-1}$ (Ho et al. 1997; Barth et al. 1999).
An ADAF cannot provide the
corresponding amount of (mainly UV) ionizing photons. This points
to a flow with a small sonic radius $r_{\rm s}$,
consistent with models where
the viscosity parameter $\alpha$ is $\approxlt 10^{-2}$ ($r_{\rm s} \sim$
a few $R_{\rm g}$; Narayan et al. 1997). Outside this radius the
flow would smoothly connect to a standard thin accretion disk.

The BeppoSAX data alone cannot rule out the possibility that a substantial
contribution to the hard X-ray emission comes from a nuclear relativistic
jet. {\it Chandra} will have enough spatial resolution
to image X-ray structures on scales as small
as $\sim$100~pc. However, it is to be noted that the ROSAT HRI images
of NGC~1052 unveiled only a soft X-ray extended emission on a much wider scale
($\sim$15~kpc) than that of the radio structures (G99), 
therefore probably due to the
emission of diffuse intergalactic gas. The contribution of this component
to the hard X-ray emission is most likely negligible.

\begin{acknowledgements}
  
The BeppoSAX satellite is a joint Italian-Dutch program.This research has made
use of the NASA/IPAC Extragalactic Database (NED) which is operated by the Jet
Propulsion Laboratory, California Institute of Technology, under contract with
the National Aeronautics and Space Administration. 

\end{acknowledgements}


\begin{thebibliography}{}

\bibitem{} Allen S.W., Di Matteo T., Fabian A.C., 2000, MNRAS 311, 493

\bibitem{} Antonucci R., Miller J., 1985, ApJ 297, 621

\bibitem{} Barth A.J., Filippenko A.V., Moran E.C., 1999, ApJ 525, 673

\bibitem{} Becklin E.E., Tokunaga A.T., Wynn-Williams C.G., 1982, ApJ 263, 624

\bibitem{} Boella G., Chiappetti L., Conti G., et al., 1997, A\&AS 122, 372

\bibitem{} Cagnoni I. della Ceca R., Maccacaro T., 1998, ApJ 493, 54

\bibitem{} Di Matteo T., Quataert E., Allen S.W., Narayan R., Fabian A.C., 2000, MNRAS 311, 507

\bibitem{} Elvis M., Wilkes B.J., McDowell J.C., et al., 1994, ApJS 95, 1

\bibitem{} Fabian A.C., Rees M.J., 1995, MNRAS, 277, L55

\bibitem{} George I.M., Fabian A.C., 1991, MNRAS, 249, 352

\bibitem{} Ghisellini G., Haardt F., Matt G., 1994, MNRAS, 267, 743

\bibitem{} Guainazzi M., Antonelli L.A., 1999, MNRAS 304, L15 (G99)

\bibitem{} Guainazzi M., Matteuzzi A., 1997, SDC-TR-014, available at {\tt ftp://www.sdc.asi.it}

\bibitem{} Guainazzi M., Perola G.C., Matt G., et al., 1999, A\&A 346, 407

\bibitem{} Heckman T.M., 1980, A\&A 87, 152

\bibitem{} Ho L.C., Filippenko A.V., Sargent W.L.W., 1997, ApJS 112, 315

\bibitem{} Krolik J.H., Madau P., Zycki P.Y., 1994, ApJ 420, 57

\bibitem{} Leahy D.A., Creighton J., 1993, MNRAS 263, 314

\bibitem{} Magdziarz P., Zdziarski A.A., 1995, MNRAS 273, 837

\bibitem{} Magorrian J., Tremaine S., Richstone D., et al., 1998, AJ, 115, 2285

\bibitem{} Malaguti G., Bassani L., Cappi M., et al., 1999, A\&A 342, L41

\bibitem{} Mass-Hesse J.M., Rodr\'iguez-Pascal P.M., Cordoba L.S.F., et al., 1995, A\&A 298, 22

\bibitem{} Matt G., 2000, Proceedings of the Conference ``X-ray Astronomy '999'', Malaguti G., Palumbo G. \& White N. eds., (Gordon \& Breach:Singapore), in press (astroph/0007050)

\bibitem{} Matt G., Perola G.C., Piro L., Stella L., 1992, A\&A 263, 453

\bibitem{} Matt G., Brandt W.N., Fabian A.C., 1996, MNRAS 280, 823

\bibitem{} Nandra K., George I.M., Mushotzky R.F., Turner T.J., Yaqoob T., 1997, ApJ 476, 70

\bibitem{} Narayan R., Kato S., Honma F., 1997, ApJ 476, 49

\bibitem{} Narayan R., Yi I., 1995, ApJ 452, 710

\bibitem{} Netzer H., Turner T.J., 1997, ApJ 488, 694

\bibitem{} Parmar A.N., Martin D.D.E., Bavdaz M., et al., 1997, A\&AS 122, 309

\bibitem{} Rees M.J., Phinney E.S., Begelman M.C., Blandford R.D., 1982, Nat 295, 17

\bibitem{} Reeves J., Turner M., 2000, MNRAS, in press (astroph/0003080)

\bibitem{} Reynolds C.S., 1997, MNRAS 286, 513

\bibitem{} Turner T.J., George I.M., Nandra K., Mushotzky R.F., 1997, ApJS 113, 23

\bibitem{} Weaver K.A., Wilson A.S., Henkel C., Braatz J.A., 1999, ApJ 520, 130

\bibitem{} Wilkes B.J., Elvis M., 1987, ApJ 323, 243

\end{thebibliography}
\end{document}